\documentclass[%
 reprint,
 amsmath,amssymb,
 aps,
 prb,
]{revtex4-2}

\usepackage{graphicx}% Include figure files
\usepackage{comment}
\graphicspath{{figures/}}
\usepackage{dcolumn}% Align table columns on decimal point
\usepackage{bm}% bold math
\usepackage{hyperref}% add hypertext capabilities
\usepackage{color}

\newcommand{\hili}{}

\begin{document}

\preprint{APS/123-QED}

\title{Fine-tuning of universal machine-learning interatomic potentials for high-entropy alloys with application to 2D (Mo,Ta,Nb,W,V)S$_2$}
%\title{Fine-tuning of universal machine-learning interatomic potentials for 2D (Mo,Ta,Nb,W,V)S$_2$ high-entropy alloys}
%\title{Fine-tuning of universal machine-learning interatomic potentials for 2D transition metal dichalcogenide alloys}
% Force line breaks with \\
%\thanks{A footnote to the article title}%

\author{Chun Zhou}
\affiliation{%
Microelectronics Research Unit, Faculty of Information Technology and Electrical Engineering,
University of Oulu, P.O. Box 4500, Oulu, FIN-90014, Finland
}%
\author{Hannu-Pekka Komsa}
\email{hannu-pekka.komsa@oulu.fi}
\affiliation{%
Microelectronics Research Unit, Faculty of Information Technology and Electrical Engineering,
University of Oulu, P.O. Box 4500, Oulu, FIN-90014, Finland
}%

\date{\today}% It is always \today, today,
             %  but any date may be explicitly specified

\begin{abstract}
High-entropy alloy (HEA) {\hili materials} and their two-dimensional counterparts (2D-HEAs) have recently attracted attention due to their tunable properties and catalytic potential, yet their chemical complexity makes direct density functional theory (DFT) calculations computationally prohibitive. The complexity also makes training of machine-learning interatomic potentials (MLIPs) challenging, but this could possibly be overcome by employing universal MLIPs as starting point.
In this work, we investigate the applicability of universal MLIP models for 2D transition metal dichalcogenide HEAs and develop effective fine-tuning strategies. 
Training structures are systematically generated and selected, and the performance of universal and fine-tuned models are benchmarked against DFT. 
We find that all universal MLIPs employed in this work yield unsatisfactory mixing energies without fine-tuning.
Applied to the experimentally synthesized (Mo,Ta,Nb,W,V)S$_2$ system, fine-tuned models based on enumerated structures can achieve near-DFT accuracy in predicting mixing energies while enabling Monte-Carlo simulations and random structure sampling at scales inaccessible to DFT. 
\end{abstract}

%Use showkeys class option if keyword
                              %display desired
\maketitle

%\tableofcontents

\section{Introduction}

The concept of high-entropy alloys (HEAs) has significantly expanded the compositional landscape in materials science, since their first inception in 2004\cite{cantor2004microstructural,yeh2004nanostructured}. By combining multiple principal elements in (near-)equimolar ratios, the enhancement of configurational entropy leads to stabilization of single-phase solid solutions with new structural and functional properties. During the past two decades, this concept has been successfully extended beyond metallic alloys to oxides, carbides, borides, nitrides, etc.\cite{rost2015entropy,yan2018high,gild2016high,gild2018high,oses2020high}, and more recently, to two-dimensional (2D) materials \cite{divilov2024disordered,ying2021high,nemani2023functional}. 
{\hili In these cases, the term high-entropy alloy refers specifically to substitutional disorder on the transition-metal sublattice of the TMDC structure, while the O/C/B/N/S sublattice remains chemically ordered.}
%Various physical properties of 2D HEAs have been investigated, such as superconductivity, magnetic ordering, and metal–insulator transitions\cite{ying2021high,chen2022insulator}. 
By integrating the high specific surface area of two-dimensional materials with the tunability and stability of high-entropy materials, 2D HEAs offer abundant and designable active sites, rendering them promising candidates for efficient, stable and multifunctional catalytic platforms, such as for hydrogen evolution reaction (HER) and carbon dioxide electrocatalysis\cite{wang2022two,cavin20212d,akhound2025activating}. Its layered hydroxides counterparts also exhibit excellent oxygen evolution reaction (OER) activity\cite{liu2024manipulating,wu2023high,li2022bottom}. 
%Specifically, the relationship between the catalytic activity of 2D transition metal dichalcogenides(TMDCs) HEAs and the ground-state phases (1T, 2H) of their constituent TMDCs has been thoroughly studied and explained\cite{akhound2025activating}.
In particular, the transition-metal dichalcogenide (TMDC) system (Mo,Ta,Nb,W,V)S$_2$ investigated in this work has garnered considerable attention, with extensive studies focusing on its synthesis routes\cite{tanaka2023growth}, catalytic performance\cite{cavin20212d}, and tribological characteristics\cite{wang2025self}.
%{\hili Here, the term high-entropy alloy refers specifically to substitutional disorder on the transition-metal sublattice of the TMDC structure, 
%In the studied 2D (Mo,Ta,Nb,W,V)S$_2$ system, Mo, Ta, Nb, W, and V occupy the metal sublattice in near-equimolar concentrations, 
%while the sulfur sublattice remains chemically ordered. Therefore, the configurational entropy relevant to the HEA description is associated with the metal sublattice.}

% P2: Importance of modeling of HEAs
% Challenges in modeling in HEAs
Although HEAs and 2D HEAs hold a great promise, their multicomponent composition and chemical disorder continue to hinder comprehensive exploration and realization of their full potential.
Density functional theory (DFT) has been considered a standard tool for exploring structural stability, electronic properties, catalytic activity, etc. However, several fundamental obstacles limit its direct application to high-entropy systems. First, the inherent chemical disorder requires large supercells and/or ensemble averaging to capture local chemical order and short-range correlations, making computations prohibitively expensive. Second, the vast configurational and compositional spaces complicate comprehensive sampling of the relevant states. 
These problems become particularly pertinent when studying temperature-dependent evolution of structural short- and long-range order, including phase segregation, via Monte Carlo simulations.
%Third, properties such as diffusion, segregation, and phase stability require access to long timescales and system sizes far beyond the reach of conventional ab initio molecular dynamics\cite{liu2023machine,jmi.2022.41}. 
In the case of 2D HEAs, interaction with the environment, whether other 2D layers, substrate, protective layers or atmosphere, can further complicate the modeling \cite{nemani2023functional}.

With the rapid development of machine learning interatomic potentials (MLIP) over the past two decades\cite{behler2007generalized,bartok2010gaussian,shapeev2016moment,mishin2021machine}, research approaches combining ML methods with suitable descriptors have been widely applied to the study of metal HEAs\cite{byggmastar2021modeling,zhang2020robust,liu2021monte,zhang2020phase,dai2020using,pei2020machine,liu2025anisotropic}. In recent years, significant research efforts have been devoted to the development of universal MLIPs (uMLIPs), such as CHGNet\cite{deng_2023_chgnet}, MACE\cite{batatia2022mace}, MatterSim\cite{yang2024mattersim}, and Allegro\cite{musaelian2023learning}.
%has further advanced the study of high-entropy systems. 
%{\hili Recent studies have further reviewed computational and MLIP-based descriptions of short-range order in HEAs and demonstrated machine-learning-potential molecular dynamics for refractory HEAs under extreme loading conditions~\cite{liu2025computational, liu2025anisotropic}.}
%
In contrast to traditional ML+descriptor methods that rely on hand-crafted atomic descriptors, uMLIPs are usually trained using large-scale databases (e.g., MPtrj\cite{deng2023materials}) to learn effective descriptors and should in principle be applicable to any material system.
Their capability to accurately describe surface energies, formation energies, and vibrational properties has been systematically validated against DFT calculations \cite{focassio2024performance,yu2024systematic,Berger25_Small}. 
%These hold tremendous potential for study of HEAs. 
While in complex systems, such as elemental alloys, uMLIPs often yield unsatisfactory predictions\cite{li2024extendable}, fine-tuning can markedly improve their accuracy without compromising transferability\cite{liu2025fine,radova2025fine}.
The selection of an appropriate training set is undoubtedly critical in such cases, whether a model is trained from scratch or by fine-tuning. Previous selection strategies include active learning \cite{rao2022machine} and Automated Small SYmmetric Structure Training (ASSYST) \cite{poul2025automated}, which enables an efficient generation of suitable training structures by exhaustively sampling within symmetry groups and stoichiometries. 
In both cases, the selected structures were used to train models from scratch. However, to the best of our knowledge, %up to September 2025, 
there have been no reports on investigating fine-tuning strategies for uMLIPs in the context of high-entropy alloy systems.
%Notably, some challenges have been addressed. For example, MLIPs trained with active-learning strategies have captured short-range order and lattice distortions in multicomponent alloys with reasonable accuracy, and machine-learning models have accelerated adsorption-energy predictions for high-entropy catalysts. The vibrational properties of 2D HEAs Mo$_{1-x}$W$_x$S$_{2-2y}$Se$_2y$ have been explored using MLIPs\cite{siddiqui2024machine}. Descriptor-based approaches, such as the disordered enthalpy–entropy descriptor (DEED), have improved the computational screening of high-entropy ceramics\cite{divilov2024disordered}. However, there is still no method available for calculating the total energy and formation energy of 2D HEAs that is both fast and accurate, representing a significant gap in the field. In particular, the latter is critical, as the enthalpy of HEAs plays an essential role in analyzing their phase transformation processes.

In this work, we benchmark a set of leading uMLIPs for predicting the energies and mixing energies of 2D HEA system (Mo,Ta,Nb,W,V)S$_2$ and investigate effective fine-tuning strategies.
%the DFT calculations of the five TMDCs (MoS$_2$, TaS$_2$, NbS$_2$, WS$_2$, VS$_2$) involved in the alloying and the corresponding MPtrj data used to train uMLIP. 
%We demonstrated that employing enumerated structures as the training set allows for more efficient fine-tuning of the model than using randomly generated structures. Specifically, we describe the generation and selection of training structures, benchmark the accuracy of both pretrained and fine-tuned models against DFT, and assess the transferability of a given model across different structural configurations. 
We consider training sets consisting of either random structures or structures generated via enumeration, and three different benchmark databases covering different types of alloy compositions.
We carefully analyze the balance between the dataset size and the model accuracy.
Finally, we demonstrate the efficacy of the fine-tuned models by applying it to the investigation of phase decomposition behavior of (Mo,Ta,Nb,W,V)S$_2$ by Monte Carlo simulations and random structure sampling.

\section{Methods}

All DFT calculations were performed using the projector augmented-wave (PAW)\cite{blochl1994projector} method implemented in the VASP package\cite{hafner2008ab}, with the Perdew–Burke–Ernzerhof (PBE)\cite{perdew1996generalized} exchange-correlation functional to treat the interactions between atoms.
We paid special attention to guarantee that our fine-tuning data is consistent with the data used to train the foundation models, i.e., employing Materials Project compliant parameters.
The \_pv pseudopotentials were used for all metal elements and standard potential for S, the plane-wave energy cutoff was set to 520 eV. For the primitive cells, a $\Gamma$-centered $12 \times 12 \times 1$ k-point mesh was employed, and comparable k-point sampling densities were applied to the supercells and special quasi-random structures (SQSs). 
All TMDCs were modeled in their 2H phase.

%Details of foundation model calculations.
%,batatia2025design
We employed several uMLIPs, including MACE-small, MACE-small-0b2\cite{batatia2022mace,batatia2025design}, MatterSim-1m, MatterSim-5m\cite{yang2024mattersim}, and CHGNet\cite{deng_2023_chgnet}. 
%The corresponding Python scripts used to run these models are provided in the supplementary materials. 
While newer and potentially better-performing uMLIPs have been recently released, the fine-tuning strategies examined in this work should be equally valid for all modern uMLIPs. In order to understand if these models have already seen TMDC alloy structures, we screened the MPTrj dataset which was used to train CHGNet and MACE models \cite{deng2023materials}.
As summarized in the Supplementary Materials Section I,  MPTrj contains all endpoint structures and a few small supercell alloy structures for four out of ten possible binary compounds, but no data for a larger number of elements. Exact details of the MatterSim training dataset are not publicly available and thus could not be verified.
%These models were selected because, when this work was started, they were among the few publicly available uMLIPs with pretrained checkpoints and direct compatibility with the Atomic Simulation Environment (ASE)\cite{larsen2017atomic}, allowing all models to be evaluated within the same workflow. As summarized in Supplementary Table~\ref{tab:foundation-domain}, their training datasets provide relevant elemental and multicomponent chemical coverage, but exact overlap with the specific 2D high-entropy TMDC configurations studied here cannot be verified from public metadata. This partial domain overlap further motivates system-specific fine-tuning.

%Details of fine-tuning.
%For comparison with DFT calculations, we used a set of 130 equimolar enumerated structures ranging in size from 2 to 10 primitive units, and 40 equimolar random structures with 4×4 supercells as show in Fig.~\ref{fig:enumeration}.
During the fine-tuning process, all configurations in relaxation trajectories have been employed as the training set, 80\% of the configurations were randomly selected for training, 15\% for testing, and 5\% for validation. The loss function was weighted with a ratio of 100:10:1 for energy, forces, and stress, respectively. The batch size and number of epochs were adjusted according to the size of the training set. The function was optimized using the AMSGrad variant of Adam\cite{kingma2014adam}. 
Upon completion of the training, almost all fine-tuned models for multi-component alloys reached a high level of convergence, with root-mean-square errors (RMSEs) on the validation set reaching $\sim$0.2 meV for energy and $\sim$5.0 meV/Å for forces. Comparable error levels were observed on the test set, indicating consistent model generalization performance.

%Details of models, such as enumerated structure generation using icet

The enumerated structures used for model training were generated using the Interaction Cluster Expansion Toolkit (ICET)\cite{aangqvist2019icet} which employs enumeration algorithms proposed by Hart et al.~ \cite{hart2008algorithm,hart2009generating}. 
%For 2D materials, periodic boundary conditions were redefined accordingly to reflect their reduced dimensionality. 
The generation of special quasirandom structures (SQS)\cite{zunger1990special, van2013efficient} and Monte Carlo annealed configurations for training purposes was also performed using tools provided within the ICET package.

% Monte Carlo simulations were carried out in combination with the fine-tuned model on a $10 \times 10$ supercell with 5000 trials 4 times for each temperature. After each trial, the resulting structure was recorded and its cluster vector (CV) extracted. The mean CV was evaluated from the last 3000 trials at each temperature.
% % Cluster vector
% The definition of cluster vector (CV) for atom pairs are the following\cite{ekborg2024construction}:
% \begin{equation}
%     \mathrm{CV_{AB} = \frac{N_{AA} + N_{BB} - N_{AB}}{N_{total}}}
% \end{equation}
% \begin{equation}
%     \mathrm{CV_{AA} = \frac{N_{AA}}{N_{total}}-\frac{1}{n^2}}
% \end{equation}
% Here, $N_{\text{AA}}$ and $N_{\text{BB}}$ denote the number of pairs of like-atoms, $N_{\text{AB}}$ represents the number of pairs of unlike-atoms, $N_{\text{total}}$ is the total number of pairs, and $n$ denotes the number of metal elements in the alloy (here $n=5$).
% {\hili The sign of CV has different meanings for heteroatomic and homoatomic pairs because of the definitions used here. For heteroatomic A--B pairs, a larger CV$_{AB}$ means a lower fraction of A--B bonds. For homoatomic A--A pairs, the term $1/n^2$ in CV$_{AA}$ is the expected A--A bond fraction in an ideally random equimolar structure; therefore, a positive CV$_{AA}$ means that A--A bonds are enriched relative to random mixing.}
Monte Carlo simulations were carried out with the fine-tuned model on a $10 \times 10 \times 1$ supercell. Benchmark runs on larger supercells were found to yield very similar results (Supplementary Materials Fig.~{\hili S8}). For each temperature, four independent MC runs with 5000 trials were performed. After each trial, the resulting structure was recorded and the first-nearest-neighbor short-range order (SRO) parameters were calculated. The mean SRO values were evaluated from the last 3000 trials of each run and then averaged over the four independent runs. 

The SRO parameter was calculated using the conventional Warren--Cowley definition \cite{cowley1950approximate}, with the multicomponent form \cite{de1971number}:
\begin{equation}
      \alpha_{BC} =
      \frac{p_{BC} - X_C}{\delta_{BC} - X_C}.
\end{equation}
Here, $p_{BC}$ is the probability of finding a $C$ atom in the nearest-neighbor shell around a $B$ atom, $X_C$ is the global concentration of element $C$, and $\delta_{BC}$ is the Kronecker delta. 
%In this work, we focus on the first-nearest-neighbor shell ($m=1$). 
For the equimolar five-component alloys studied here, $X_C=0.2$ for all metal species.
With this convention, $\alpha=0$ corresponds to a random equimolar distribution. For unlike pairs ($B \ne C$), negative SRO values indicate enrichment of $B$--$C$ pairs, while positive values indicate depletion. For like pairs ($B=C$), positive SRO values indicate enrichment of homoatomic pairs relative to random mixing.

\section{Results}

\subsection{Benchmarking foundation models}

\begin{figure*}
\includegraphics[width=0.8\textwidth]{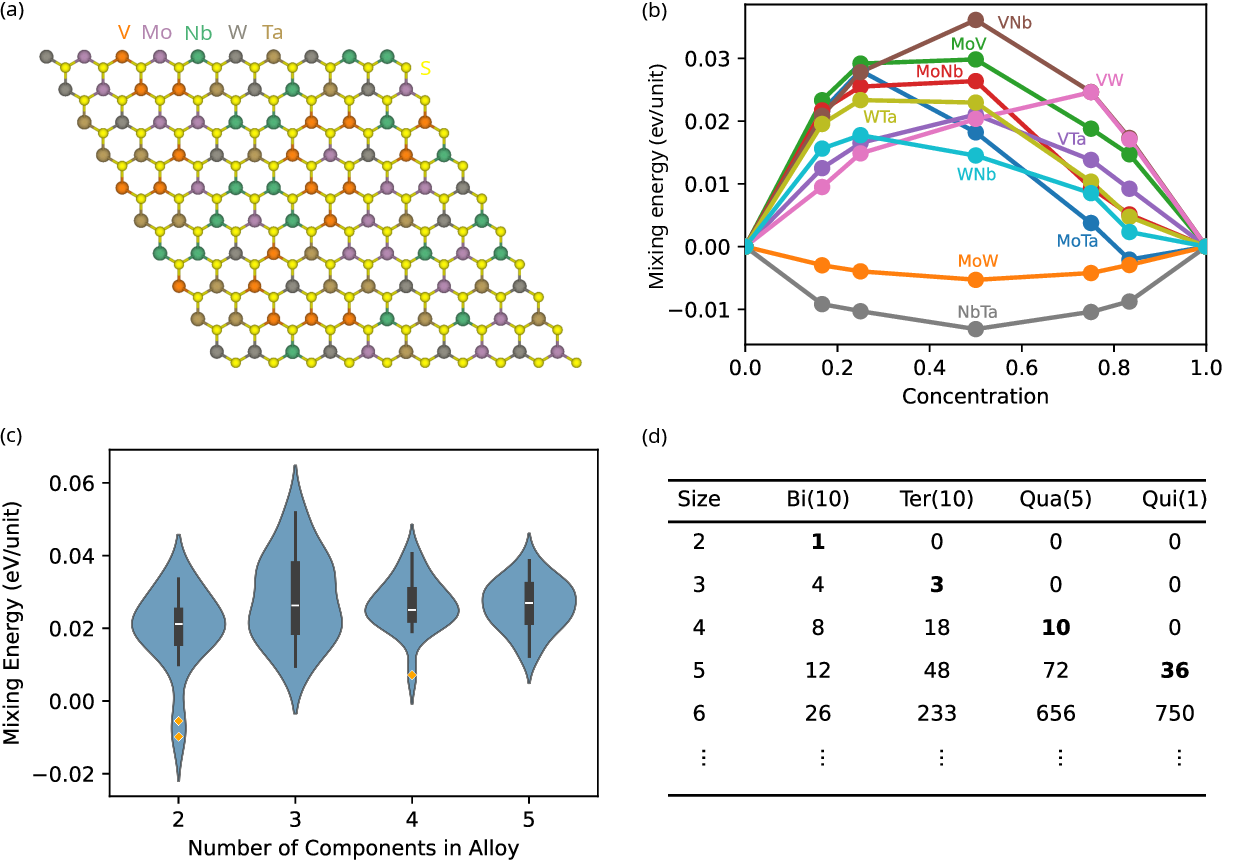}% Here is how to import EPS art
\caption{\label{fig:heas}
Illustration of TMDC HEA structure and energetics.
(a) Atomic structure of a random $10 \times 10$ supercell of (Mo,Ta,Nb,W,V)$_{0.2}$S$_2$.
(b) The mixing energy as a function of concentration for all (10) possible binary alloys. 
(c) Mixing energy of 2-, 3-, 4-, and 5-component systems with random concentrations and configurations. The white dots and orange diamonds indicate the median values and outliers, respectively. The thick black bars represent the interquartile ranges (IQR), and the thin black lines denote the whiskers, corresponding to $1.5 \times \mathrm{IQR}$.
(d) Table of the number of enumerated structures as a function of the supercell size (number of unit cells in the supercell) and the number of alloy components. The values are for single combination of elements and the numbers in parentheses refer to the number of possible element combinations [e.g., there are 10 binary combinations, as shown in panel (b)].
Bold values on the diagonal indicate equimolar structures with minimal supercell size for n-component alloy.
}
\end{figure*}

\begin{comment}
[Always first explain the point of the figure and some main things one can see from it.]
1(b-c)
Discuss the mixing energy ranges, over full concentration in the case of binaries and selected random concentrations for others.
Discuss which elements are easy to mix, which are difficult.
1(d)
Explain the difficulty in benchmarking MLIPs due to the very large composition space
(and different needs of different researchers). 
Thus we select have 3 datasets that will be used.
Describe their details.
\end{comment}

Our work focuses on the experimentally synthesized 2D (Mo,Ta,Nb,W,V)S$_2$ TMDC system \cite{cavin20212d}, whose atomic structure is illustrated in Fig.~\ref{fig:heas}(a).

To comprehensively assess the accuracy of the foundation and fine-tuned MLIP models and to mimic the diverse needs in the research of HEAs, we have selected three distinct evaluation databases:
\begin{itemize}
    \item \textbf{Database 1:} Consists of data points from all (ten) binary alloys at five different concentrations, with three randomly selected structures for each concentration, resulting in a total of 150 ($6 \times 6$) supercell structures.
    \item \textbf{Database 2:} Includes 20 structures with random concentrations and configurations for 2-, 3-, 4-, and 5-component alloys, yielding a total of 80 ($4 \times 4$) supercell structures.
    \item \textbf{Database 3:} Contains enumerated equimolar structures for alloys of N components up to size N ($N \leq 5$).
\end{itemize}

Mixing energies corresponding to Database 1 are shown in Fig.~\ref{fig:heas}(b). Among all binary alloys, the lowest mixing energy is \(-14.6\) meV/unit, observed in the equimolar (Nb,Ta)S$_2$, while the highest is \(37.7\) meV/unit, found in the equimolar (V,Nb)S$_2$. 
The isoelectronic binary alloys (Mo,W)S$_2$ and (Ta,Nb)S$_2$ have negative mixing energy through all concentrations, indicating thermodynamic miscibility at 0 K. Some of the alloys show asymmetric mixing energy curves, which may indicate contributions to lattice (in)stability arising from formation of charge density wave, magnetism, or 1T-phase near the end-point.
%nonideal solution behavior due to 
%and the potential for phase segregation. 

Fig.~\ref{fig:heas}(c) displays violin plots of the mixing energy distributions for binary, ternary, quaternary, and quinary alloy systems, corresponding to Database 2. 
The median mixing energy across all groups are comparable, but the spread narrows as the number of components increases and the mixing energy per unit gets averaged over many possible element pairs.
%On the one hand, this is due to the absence of negative mixing energies in HEAs. 
%On the other hand, for the same supercell size, alloys with more components possess a relatively larger structural pool. When sampling the same number of structures from these pools, the representativeness of the samples inevitably decreases as the size of the structural pool increases.
That said, there are also multicomponent structures with higher mixing energies than any of the binaries.
For instance, the highest mixing energy of 52 meV/unit is obtained for one of the (Mo,Nb,V)S$_2$ ternary alloy structures. It contains the Mo–Nb, Mo–V, and Nb–V pairs, which correspond to the three highest-mixing-energy pairs at a concentration of 0.5 in Fig.~\ref{fig:heas}(b). 
For quaternary and quinary alloys, the introduction of Mo–W and Nb–Ta pairs necessarily leads to a reduction in the maximum mixing energy.

Finally, Fig.~\ref{fig:heas}(d) presents a table of the numbers of enumerated structures according to the number of components and cell size.
%The color intensity indicates the total number of structures, with darker shades representing higher counts. Each small square in the grid shows two values: the number in the top-right triangle corresponds to the number of structures for a single alloy species, while the number in the bottom-left triangle represents the total number of structures for all alloys at a given composition. 
A clear trend is observed: the number of enumerated structures increases rapidly with cell size, and alloys with more components exhibit a higher rate of growth. Notably, the diagonal elements correspond to equimolar compositions covering all alloys at any number of components, making them a representative subset of the enumerated structures and used to construct Database 3.

\begin{figure*}
\includegraphics[width=0.8\textwidth]{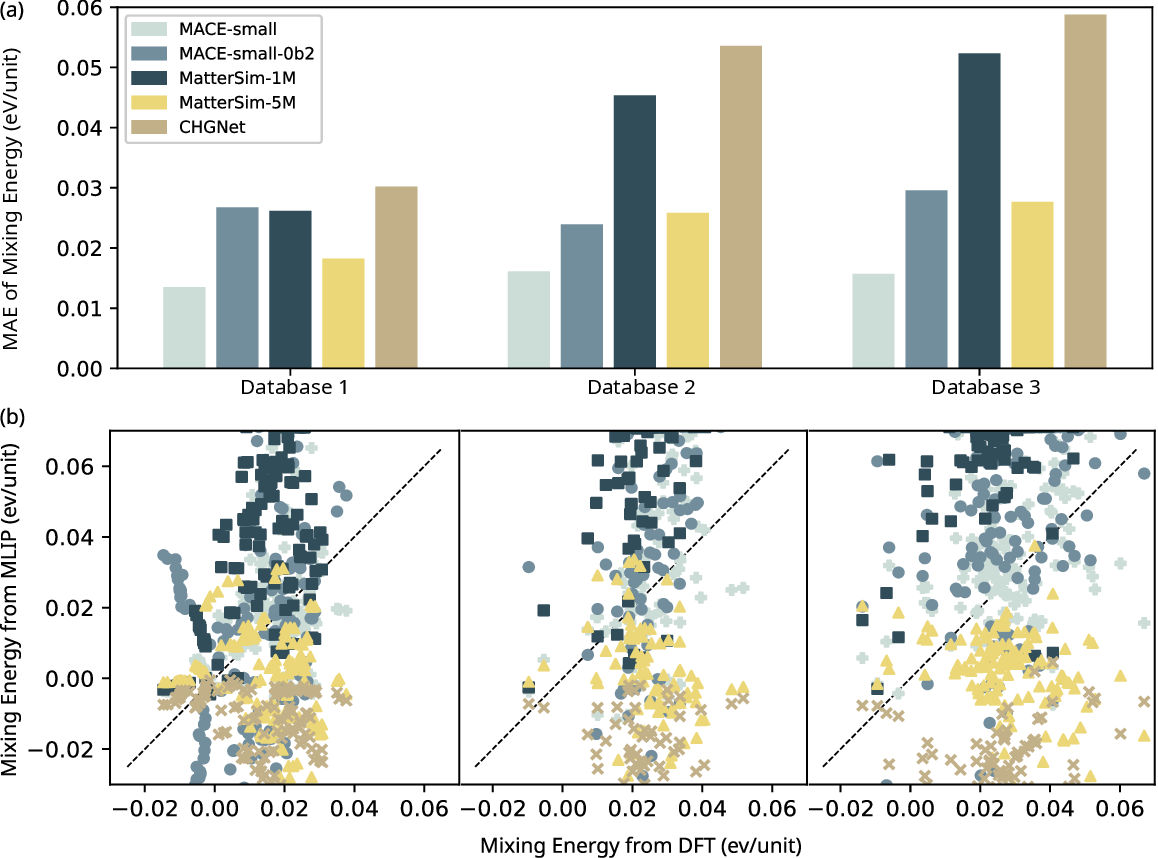}% Here is how to import EPS art
\caption{\label{fig:foundation}
uMLIPs (MACE, MatterSim, CHGNet) applied to benchmark databases.
(a) mixing energy for all three databases and five foundation models.(b) The corresponding correlation plots of mixing energy.
}
\end{figure*}

%2(a-b). Highlight that none of the UMLIPs is sufficiently good and they might fail in different ways. 

For benchmarking available universal machine-learned interatomic potentials, we selected five commonly used potentials and applied them to these three databases. As shown in the bar chart for mixing energy MAE in Fig.~\ref{fig:foundation}(a), all five uMLIPs exhibit similar performance trends across the three databases, generally showing a small increase in the error when going from Database 1 to Database 2 to Database 3.
For all databases, the MACE-based uMLIPs and MatterSim-5M exhibit relatively low prediction errors. In contrast, MatterSim-1M and CHGNet show substantially larger errors, particularly for Database 2 and Database 3. Unlike Database 1, both Database 2 and Database 3 contain HEAs, which may account for the increased prediction errors in these databases. 

Figure~\ref{fig:foundation}(b) shows the correlation plot of mixing-energy predictions from the five selected uMLIPs. The mixing-energy predictions exhibit substantial scatter for all tested models. The data either show excessive dispersion, cluster within limited energy ranges, or deviate strongly from the DFT reference trend. 
On the other hand, the corresponding plots for total energy,
%In contrast to the total-energy benchmarks 
shown in the Supplementary Information Fig.~S1, fall neatly on the diagonal. This apparent discrepancy arises simply from the much reduced energy range for mixing energies as compared to total energies.
This indicates that good total-energy correlation alone is not sufficient to guarantee reliable configurational energetics. 
In addition, the force errors collected in the Supplementary Information Fig.~{\hili S3} show that, except for MatterSim-1M, all foundation models give similar force errors in the range 60--80 meV. While non-negligible, they are small enough to guarantee convergence to the correct ground state structure, which is highly beneficial in guaranteeing stable behavior for fine-tuned models in the small-data regime.
In the end, none of the selected foundation uMLIPs provides consistently accurate mixing-energy predictions for the HEA structures, highlighting the need for fine-tuning to improve performance for this chemical and configurational space.

Considering the overall performance in predicting both total energy and mixing energy across different databases, the MACE-based uMLIPs appear to be the most suitable choice for our target alloy systems.
Although MACE-small-0b2 gives a slightly larger raw energy and mixing energy MAE than MACE-small, it was selected as the foundation model for subsequent fine-tuning because it showed more robust small-data fine-tuning behavior. As summarized in Table~S2 {\hili and Fig.~S4}, matched fine-tuning runs initialized from MACE-small-0b2 consistently reached lower final validation loss and lower validation force MAE than those initialized from MACE-small. 
%Thus, the choice of MACE-small-0b2 was based on its improved convergence and force-learning behavior during fine-tuning, rather than on the raw uMLIP energy MAE alone.
%While MACE-0b2 provides slightly inferior MAE, we chose it here as it gives improved forces, and better stability in molecular dynamics and in fine-tuning.
Thus, all fine-tuned models in the following of this work were initialized from the MACE-small-0b2 foundation model. %Thus, differences among the fine-tuned models arise from the selection of training sets rather than from different pretrained architectures.

% In Figs.~\ref{fig:foundation}(c,d) we show the correlation plots for total energy and mixing energy.
% In the case of total energy correlation plots, Fig.~\ref{fig:foundation}(c), the MACE- and MatterSim-based uMLIPs show a well-aligned prediction for total energy, with nearly all data points concentrated near the diagonal, except for CHGNet, which shifts from the central diagonal. 
% However, the predictions for mixing energy,
% Fig.~\ref{fig:foundation}(d), exhibit a dramatically different picture with considerably broader scatter for all 5 uMLIPs. These models either display excessive dispersion, form dense clusters in limited regions, or significantly deviate from the diagonal. 
% These results show that benchmarking based only on total energy correlation plots is insufficient. 
% In addition, none of the selected uMLIPs are able to provide reliable predictions for the mixing energies of HEAs, highlighting the necessity of fine-tuning these models for more accurate performance.

\begin{figure*}
\centering
\includegraphics[width=0.8\textwidth]{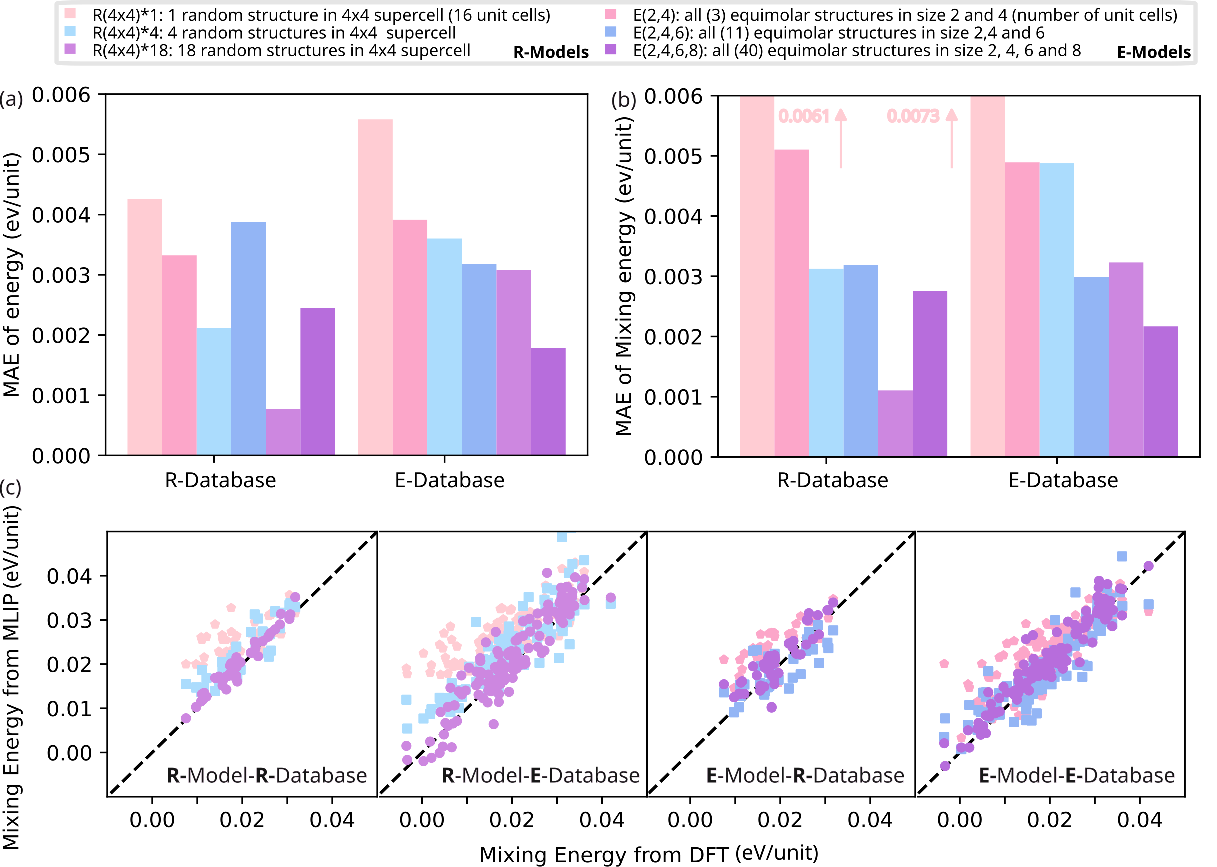}% Here is how to import EPS art
\caption{\label{fig:enumeration}
Comparison of random and enumerated models in the (Mo,Ta)S$_2$ system with gradually increasing training set size.
(a) Bar plots with MAE values of energy and (b) mixing energy for a series of models on random and enumerated databases, respectively. See text for explanation of model labeling.
(c) Correlation plots of mixing energy. 
%R/E(M) refer to random/enumerated model and R/E(D) to random/enumerated database. 
All random/enumerated models are plotted in each panel, with the same color coding as in panels (a,b).
}
\end{figure*}
%(c) Bar plot with 3 RMSE values from 3 datasets, 5 models (3 enumerated+binaries+random)
%(d) Correlation plots?

%The enumerated set covers all possible nearest-neighbor environments much more efficiently than random ones.

\subsection{Fine-tuned models for binary alloy}

A critical step in fine-tuning the models is the selection of an appropriate training set. 
It is extremely challenging to achieve comprehensive sampling of the vast chemical and configurational space of HEAs, which is one of the main reasons why in previous MLIP studies for HEAs random structures have often been employed in training set\cite{cao2025capturing,kostiuchenko2019impact}. However, for 2D structures, this difficulty is significantly reduced, and the number of enumerated structures shown in Fig.~\ref{fig:heas}(d) remains tractable.
These two approaches are compared in Fig.~\ref{fig:enumeration}, where we use the (Mo,Ta)S$_2$ system as an example to cross-validate the performance of models fine-tuned using random structures and enumerated structures, respectively, by comparing their predictions with DFT results.

For this binary test case, two reference databases were constructed for the model evaluation. The random database contains 50 randomly generated equimolar $4\times4$ (Mo,Ta)S$_2$ structures. The enumerated database contains all equimolar enumerated (Mo,Ta)S$_2$ structures with supercell sizes from 2 to 10, giving 130 structures in total (Supplementary Table S1). 
Six fine-tuned models were trained using selected structures from either the random or enumerated structure pools, three models each.
%For both the random and enumerated models, three training sets were selected. 
To ensure a fair comparison, the total number of atoms in each corresponding pair of training sets was kept approximately the same.
%{\hilifor For each selected structure, the corresponding DFT relaxation trajectory was split into training, validation, and test subsets as described in the Methods section.}
%For the enumerated models, the three training sets consist of all equimolar structures of supercell sizes in the range 2--4, 2--6, and 2--8 (Supplementary Table S1). The corresponding training sets for the random models contain 1, 4, and 18 equimolar $4 \times 4$ supercells, respectively.
The labels in Fig.~\ref{fig:enumeration} denote the training sets used for fine-tuning. Here, E and R denote enumerated and random equimolar structures, respectively. E(2,4), E(2,4,6), and E(2,4,6,8) include all equimolar enumerated structures with the listed supercell sizes. Similarly, R(4$\times$4)*n denotes a random training set containing n equimolar $4\times4$ supercells.
%The corresponding random training sets contain 1, 4, and 18 equimolar $4 \times 4$ supercells, respectively.
The dataset was split into training, test, and validation subsets as described in the Methods section. An exception is made for the smallest models in both the random and enumeration groups, which were divided only into training (90\%) and validation (10\%) subsets due to their limited dataset sizes.

Fig.~\ref{fig:enumeration}(a) and (b) show that each model consistently performs better on the type of structures it was trained on. Moreover, both the random and enumerated models exhibit a decreasing trend in MAE for both types of structures as the size of the training dataset increases. The only non-monotonic result is the E(2,4,6) model in Fig~\ref{fig:enumeration}(a), which shows a slightly higher error on the R-data compared to the E(2,4) model.
%This may be due to the non-linear coupling between the model, data, and training process, where a larger training set can sometimes lead to worse performance\cite{viering2022shape,loog2022survey}.

However, when the number of training structures is relatively small [E(2,4),E(2,4,6), R($4 \times 4$)*1, R($4 \times 4$)*4], the fact that the enumerated models achieve mixing energy MAEs on random structures comparable to or even lower than those of the random models demonstrates their higher fine-tuning efficiency. In contrast, as the number of training structures increases, the random model exhibits a more significant reduction in MAE on the random structures; however, its overall performance on the enumerated structures remains relatively poor, reflecting to some extent the limited transferability of the random model.
On the other hand, the enumerated model exhibits a smaller difference in MAE between the random and enumerated structures, especially for the mixing energy in Fig.~\ref{fig:enumeration}(b). This indicates a more stable transferability of the enumerated model, and its lower accuracy on the random structures can be compensated by expanding the training set.

The correlation plots of mixing energy, Fig.~\ref{fig:enumeration}(c), show the distributions of the random and enumerated model energies on the two types of structures. 
It is worth noting, that the enumerated structures exhibit a wider range of mixing energies than the random structures.
Enumeration samples the configurational phase space more comprehensively, including ordered structures that are expected to yield the minimum or maximum mixing energy.
Thus, although models trained using enumerated structures might show higher MAE than those trained using random structures (at least for some training datasets), they are expected to provide safer predictions due to the lack of "blind spots".
%providing an advantage for models aimed at predicting energies and mixing energies by offering more comprehensive energy sampling.
Therefore, for more complex alloy systems, we prioritize using enumerated structures for fine-tuning.

\subsection{Fine-tuned models for high-entropy alloys}

\begin{figure*}
\centering
\includegraphics[width=1\textwidth]{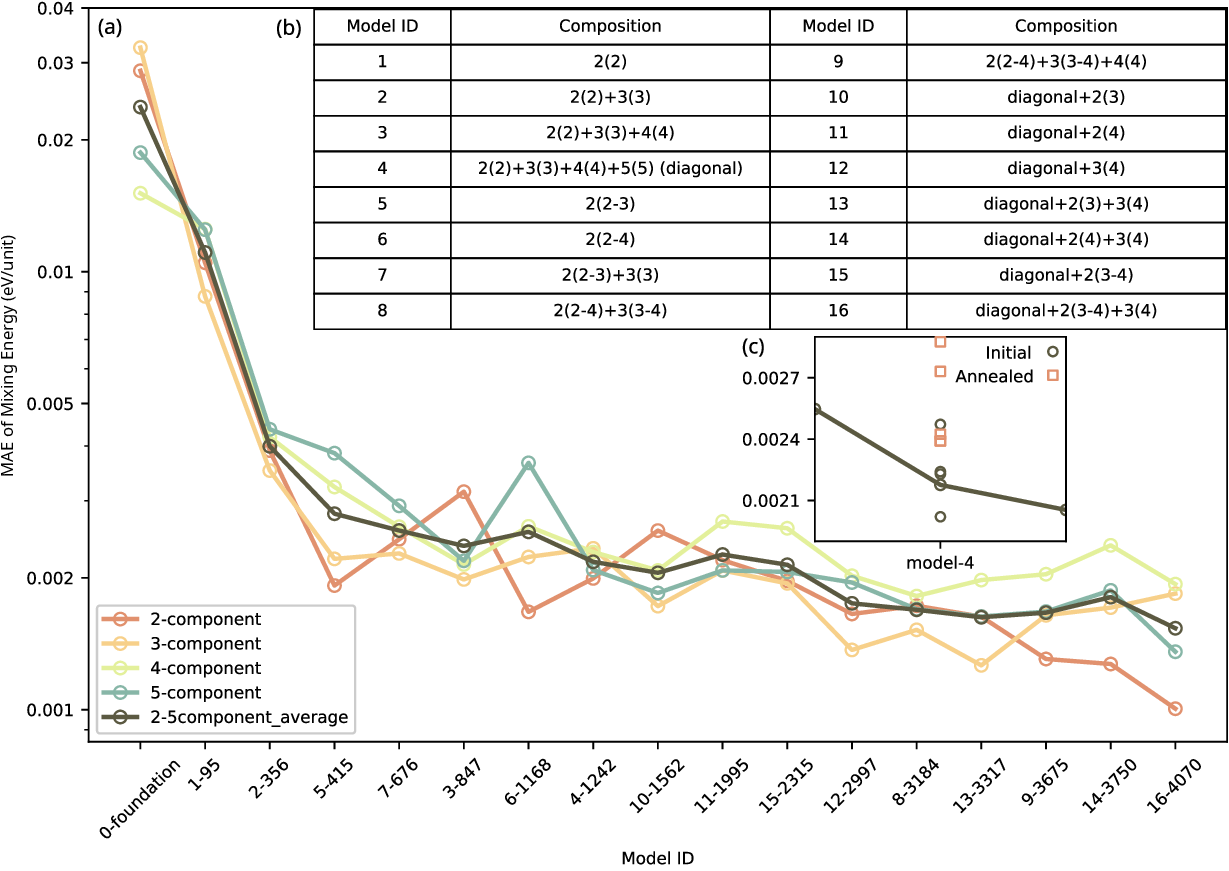}% Here is how to import EPS art
\caption{\label{fig:finetune}
(a) Foundation model and 16 fine-tuned models applied to random 2-, 3-, 4-, and 5-component alloys (Database 2). The labels on the x-axis indicate the model ID and the number of configurations in the training file. 
(b) Table describing the composition of training data used in fine-tuning these models. The numbers outside parentheses denote the number of alloy elements, while the numbers in the parentheses indicate the supercell sizes.
(c) Illustration of the stochastic effects observed in Model 4 and its Monte Carlo annealing counterpart.
}
\end{figure*}

To achieve better predictive performance of models trained on enumerated structures, instead of using only equimolar structures, we expanded the training set to include all possible structures with varying concentrations. This comprehensive inclusion enables coverage of all possible local atomic environments, which is expected to improve the predictive capability of the model\cite{li2024local}. %Model5 and E2-6 in fig3
We constructed a series of training sets from different combinations of the enumerated structures, as listed in Fig.~\ref{fig:heas}(d), and used them for fine-tuning. In total, 16 different training sets were selected, as detailed in Fig.~\ref{fig:finetune}(b), and the corresponding fine-tuned models were evaluated on Database 2, Fig.~\ref{fig:finetune}(a). In Fig.~\ref{fig:finetune}(b), "2(3)", for example, refers to all enumerated binary structures with supercell size 3, i.e., consisting of compositions 1/3 and 2/3.
The number of configurations on the x-axis refers to the number of structures selected to the training set from the relaxation trajectories, whereas the number of distinct structures is roughly one tenth of that and can be calculated for each case from Fig.~\ref{fig:heas}(d).
This setup allows us to examine how prediction performance evolves with the size of the training set, as well as how different combinations of training data, under similar data volumes, affect the model’s ability to predict alloys with varying compositions. 
We emphasize, that none of the structures in Database 2 were used in training these models and contain structures with any number of elements, from binary alloys to HEAs.

From the results in Fig.~\ref{fig:finetune}, it can be observed that, across all models, the MAEs for alloys with different compositions eventually converge to approximately $2~\mathrm{meV/unit}$, and their average converges to $1.5~\mathrm{meV/unit}$. This represents a significant improvement over the MAE of 
$25~\mathrm{meV/unit}$ of the MACE-0b2 foundation model or $16~\mathrm{meV/unit}$ of MACE-small model shown in Fig.~\ref{fig:foundation}(b). 
Due to its limited training set, Model 1 exhibits only a modest improvement in accuracy. In contrast, Model 2 shows a substantial enhancement across all cases, despite the absence of 4- and 5-component alloy structures in its training set. Building upon Model 2, Model 3 achieves further improvement overall by introducing 4-component alloy, though the gain for binary alloys is relatively minor. Models 5 and 6, whose training sets consist solely of binary alloy structures, perform best on binary alloys but their accuracy deteriorates significantly as the alloy compositions become more complex. With increasing training set size and structural diversity, the average error of the fine-tuned models across 2- to 5-component alloys is largely converged at Model 12, although the minimum is reached for the largest model, Model 16. Notably, Model 16 also achieves the lowest errors for both binary and quinary alloys.
%
%Relative to the average mixing energy of $0.025~\mathrm{eV/unit}$ of its foundation model in Database~2 as show in Fig.~1(c), the overall prediction error corresponds to approximately $6.4\%$. 
This indicates that, with sufficient and representative training data, the model can achieve high accuracy across a wide range of alloy compositions.
However, if achieving the lowest possible error is not critical, reasonably accurate models can still be obtained with relatively small training sets. For example, Model 4 achieves an accuracy close to convergence, while requiring significantly fewer training structures (126 vs 426 distinct structures in Models 4 and 16, respectively).
It should be noted that Model 4 actually includes all structures along the diagonal of Fig.~\ref{fig:heas}(d), which for HEAs represents a theoretically minimal but complete training set in the sense that it includes structures with all possible element combinations. 
%Moreover, although Model 4 is not the most accurate in terms of overall MAE, it exhibits the smallest variation in MAE across alloys with different compositions. %This indicates that Model 4 may be particularly suitable for applications requiring consistent performance across a wide range of compositions.

As mentioned in the Methods section, the dataset was randomly partitioned into training, test, and validation subsets. 
To assess the influence of stochastic effects on model training, Model 4 was subsequently trained multiple times with varying seed values, and the resulting dispersion is presented in Fig.~\ref{fig:finetune}(c). 
The MAEs from different models vary by a few meV/unit. 
Still taking Model 4 as an example, we also employed a Monte Carlo annealing module from ICET to find a subset of structures for which the condition number of the cluster vector matrix is minimized.
%anneal its training set with a specified output ratio, thereby using fewer structures for fine-tuning. 
The annealing process is also subject to stochastic effects. With a sampling ratio of $50\%$ of the total structures, multiple annealing simulations were performed, showing that roughly $80\%$ of the structures overlapped between any two runs (information can be found in the Supplementary Table S2). When structures from different annealing batches were used for training with varying seed values, the resulting distribution is shown in Fig.~\ref{fig:finetune}(c). 
%In general, whether trained on the full dataset or on annealed subsets, 
Again, stochasticity introduces only a minor influence on model accuracy, MAE varying by a few meV/unit, but the average MAE increased by about 3 meV/unit.
%Notably, models trained with annealed structures yield errors comparable to those trained on the full dataset, and under favorable stochastic conditions may even achieve superior accuracy.
Interestingly, the resulting MAE is close to that from Model 7, whose dataset size is also about 50\% of Model 4, although consisting of different types of structures.
Similarly, annealing to 25\% of structures yielded average MAE of around 4 meV/unit, again fairly close to the value from Model 2 with about 25\% dataset size compared to Model 4.
Consequently, as the structures from enumeration and Monte Carlo annealing benefit the fine-tuning equally,
it appears that the fine-tuned models are still in the underfitting regime, but also that the two methods sample configuration space equally well.
%As shown in Fig. 4(c), the model trained on annealed structures exhibits a MEA very close to that of the original model; however, it is worth noting that models trained with higher proportions of structures actually yield larger errors.

\begin{figure*}
\centering
\includegraphics[width=0.8\textwidth]{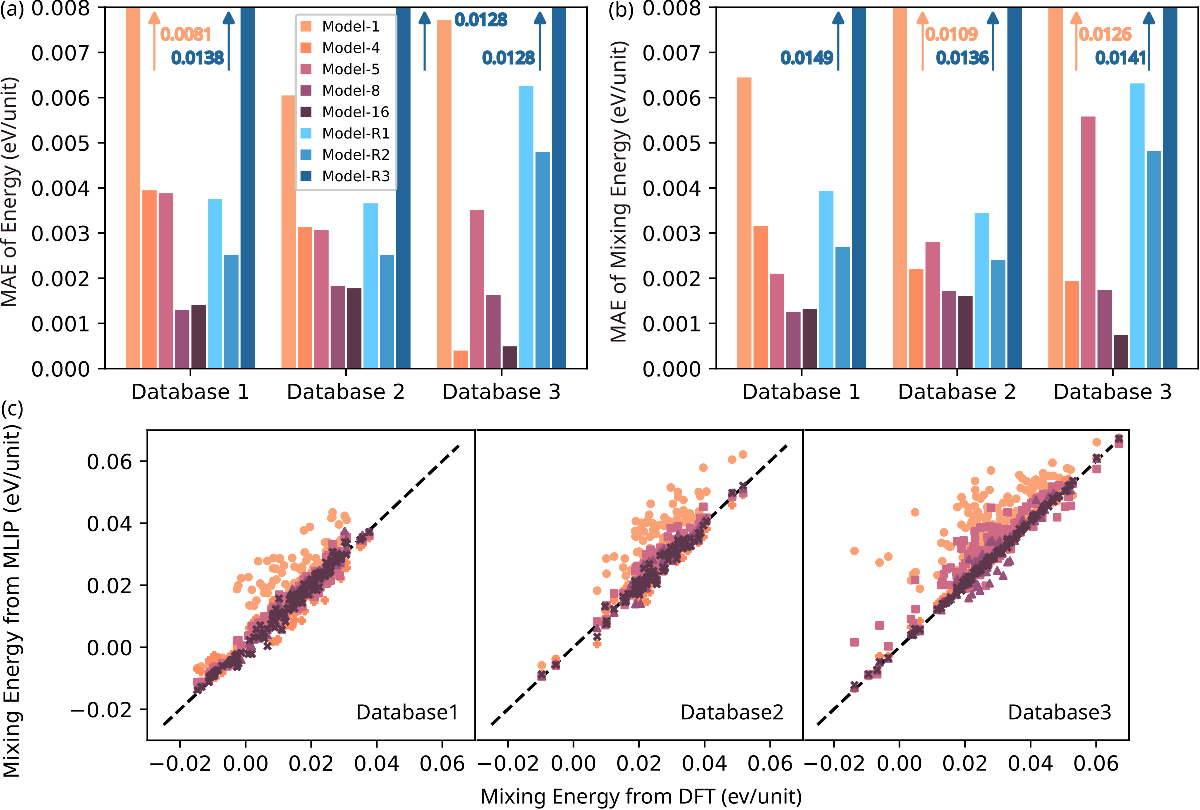}% Here is how to import EPS art
\caption{\label{fig:selected_models}
Fine-tuned models applied to databases.
(a) Bar plot with MAE values of energy and (b) mixing energy of 8 models (5 enumerated, 3 random) on 3 databases.
({\hili c }) Correlation plots of mixing energy.
}
\end{figure*}

%barplot of enumerated models
Several representative models trained on enumerated structures were selected and applied to the other two databases to further assess their performance on different types of structures, shown in Fig.~\ref{fig:selected_models}(a,b).
The selected models were observed to exhibit nearly identical error trends in Database 1 and Database 2 of both energy and mixing energy. One small difference is that in Database 1, Model 5 exhibits lower errors in mixing energy than Model 4, whereas the opposite trend is observed in Database 2. This can be attributed to the fact that the training set of Model 5 contains more binary alloy structures than Model 4 and Database 1 consists exclusively of binary alloys. In contrast, Model 4 includes all alloy types, which is responsible for its superior performance in Database 2.
Since the training structures of Models 4 and 16 include all the configurations contained in Database 3, these two models exhibit the lowest energy errors when applied to Database 3. Although Model 4 shows a larger error in the mixing energy compared to Model 16, its accuracy remains at the same level as Model 8, whose training dataset is approximately three times larger. This highlights the importance of completeness in training structures: Model 8 lacks configurations from 4- and 5-component alloys.
Model 4 performs reasonably well across all databases, despite its training set containing only 31\% of the structures of Model 16.
%correlation plots
In the corresponding correlation plots, Fig.~\ref{fig:selected_models}(c), it can also be observed that, except for Model 1, which shows considerable scatter due to its small training set, the other models are distributed close to the diagonal, with Model 16 exhibiting the most concentrated distribution.
Since Model 16 consistently exhibits (one of) the smallest prediction errors in both energy and mixing energy across the three different databases, it will be adopted in the study of phase separation below.
%it is considered the most accurate among the fine-tuned models.

%+ Model 4 good enough for most thing, at 31\% of dataset size.
%+ Model 5 good enough for everything except enumerated, at 10\% of dataset size.

% random models
To enable a comprehensive comparison between random and enumerated structures in the HEA system, three models fine-tuned using random structures were also employed. The compositions of their training sets are as follows:
\begin{itemize}
    \item \textbf{Model-R1:} $1 \times (2 \times 1) \times N_{\mathrm{Bi}}$, $1 \times (3 \times 3) \times N_{\mathrm{Ter}}$, $1 \times (6 \times 6) \times N_{\mathrm{Qua}}$, and $7 \times (2 \times 1) \times N_{\mathrm{Qui}}$ in total.
    \item \textbf{Model-R2:} All structures in Model-R1, plus $3 \times (4 \times 4) \times N_{\mathrm{Bi}}$ and $2 \times (6 \times 6) \times N_{\mathrm{Ter}}$.
    \item \textbf{Model-R3:} Quinary alloys only, containing $20 \times (5 \times 5)$ structures in total.
\end{itemize}
The seemingly complex design of Model-R1 and Model-R2 aimed to ensure that the representation of different alloy compositions and the total number of atoms closely resemble those of the enumerated Models 4 and 16, respectively. 
Model-R3 was chosen to investigate if fine-tuning with only 5-component alloy structures could yield model that performs well also for structures with lower number of components.
We also note that no structures from the three benchmark databases were used to train these three random models.

%comparison between random and enumerated models
The random counterparts of Model 4 and Model 16, Models R1 and R2 generally show lower performance than Models 4 and 16 on both energy and mixing energy across Database 1 and Database 2, except that Model R1 achieves a slightly lower energy error than Model 4 on Database 1, Fig.~\ref{fig:selected_models}(a,b)  (the corresponding correlation plots are shown in the Supplementary Fig. {\hili S5}). 
The comparison on Database 3 is unfair due to inherent bias. However, when examining the random models alone, their performance on the enumerated-structure Database 3 is the worst among the three databases.
Although Model R3 has a training set comparable in size (in terms of total number of atoms) to those of Models 4 and R1, it exhibits poor performance across all databases, performing even worse than Model 1, despite the latter having a considerably smaller training set. This may reflect the importance, for HEA systems, of including all alloy types in the training set and ensuring diversity in the sizes of the training structures.

%training from scratch
In addition to fine-tuning, we also attempted to train the models from scratch. For comparability, a model with the training set of Model 16 was first trained using parameter settings that closely match those used in fine-tuning. However, when evaluating the model on the three databases, several structures (across all databases) failed to reach a stable relaxation. Increasing the size of the model improved accuracy and allowed successful application to all databases, though at the cost of substantially reduced computational efficiency. Although the larger model yielded only slightly higher errors than its fine-tuned counterpart (see comparison in the Supplementary Fig. {\hili S6}), its slower runtime limits its suitability for e.g., large-scale Monte Carlo simulations. When Model 4 was retrained from scratch with the same increased model size but a relatively smaller training set than Model 16, the relaxations failed again on structures of Database 2. This is likely because the training set of Model 4 does not sample the configurational phase space sufficiently broadly and will fail outside its training set.
%is too small to meet the requirements of training from scratch. 
%Suggesting that training with annealed structures would be hindered by the same issue.
By optimizing the training parameters and the training set, one may get a more accurate model by training from scratch, although this possibility was not further explored in this work.
Nevertheless, our results demonstrate that fine-tuning yields stable models for all alloy configurations even when trained with a relatively small dataset, whereas those trained from scratch either need considerably large training dataset or will struggle for configurations outside the training dataset.

\begin{figure*}
\centering
\includegraphics[width=1\textwidth]{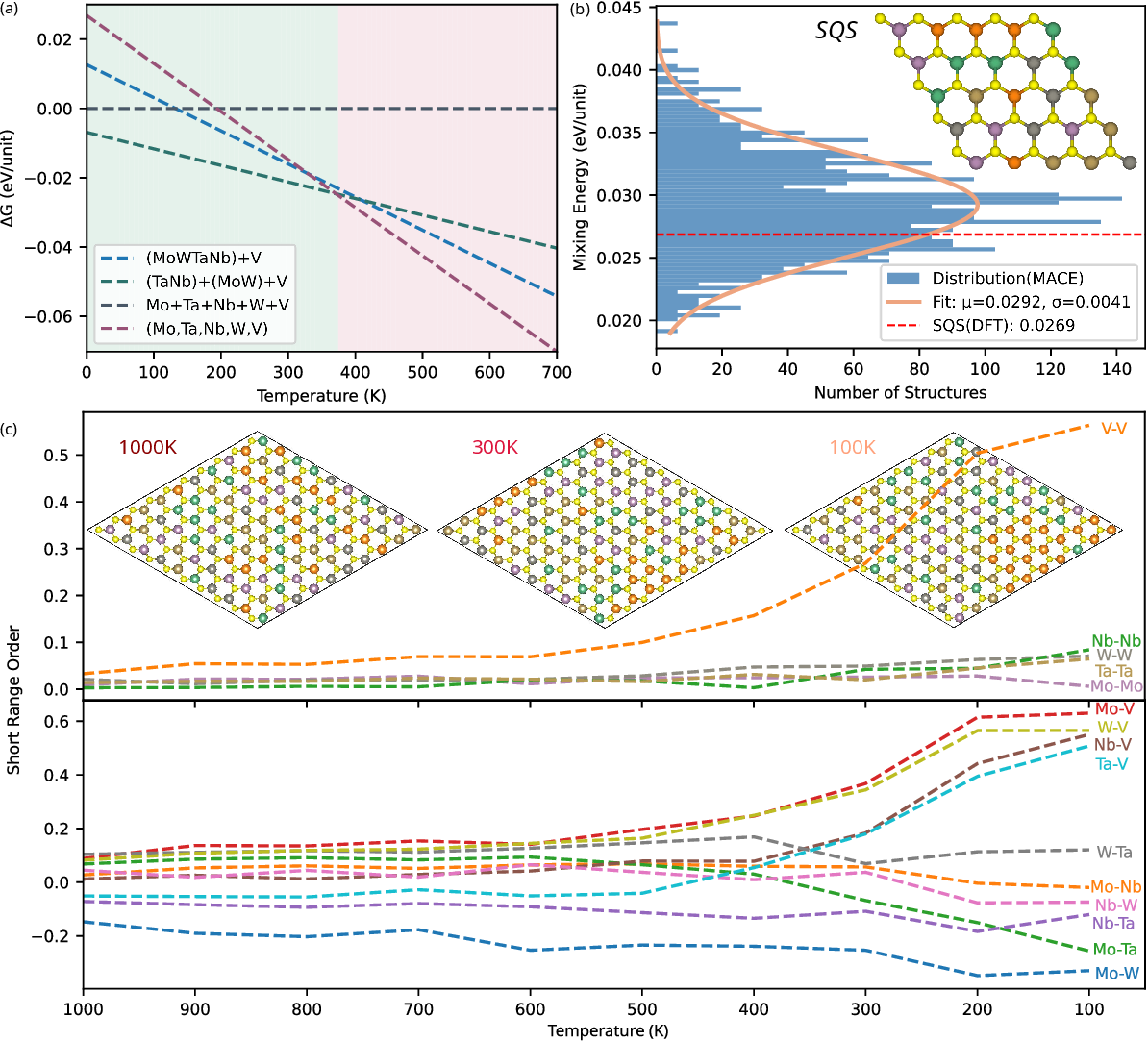}% Here is how to import EPS art
\caption{\label{fig:phase_separation}
%Monte Carlo simulations using fine-tuned model.
Phase decomposition as a function of temperature.
(a) The temperature dependence of the Gibbs free energy of the quinary alloy and its possible decompositions.
(b) The distribution of mixing energies for 500 random quinary alloy equimolar structures based on $5 \times 5$ supercells and the corresponding SQS configurations.
(c) Temperature dependence of the first-nearest-neighbor short-range order (SRO) parameters obtained from Monte Carlo simulations. For homoatomic pairs, positive SRO values indicate enrichment of like-atom bonds relative to random mixing, whereas for heteroatomic pairs, positive SRO values indicate depletion of unlike-atom bonds. Structures from selected temperatures are shown in the insets.
}
\end{figure*}

\subsection{Monte-Carlo simulations using fine-tuned model}

To analyze the phase-separation behavior of our HEA system, we used the most accurate fine-tuned model (Model 16) to study decomposition by sampling random structures and by performing Monte Carlo simulations.

%(a) and (b)
First, we studied the decomposition of the 5-component alloy by evaluating the Gibbs free energy of all (51) possible decompositions. 
The microscopic criterion for decomposition is the thermodynamic stability of a decomposed state relative to the homogeneous HEA, reflecting the competition between configurational entropy, which stabilizes chemical disorder, and enthalpic interactions, which favor chemical ordering, segregation, or phase separation in HEAs~\cite{liu2025computational}.
% For a decomposition pathway $P$, the Gibbs free-energy change is defined as
% \begin{equation}
%       \Delta G_P(T)=\sum_{i} f_{i}G_{i}(T)-G_{\rm HEA}(T),
% \end{equation}
% where $f_{i}$ and $G_{i}$ are the phase fraction and Gibbs free energy of product phase $i$, and $G_{\rm HEA}$ is the Gibbs free energy of the homogeneous alloy at the same overall composition. Thus, $\Delta G_P<0$ indicates a thermodynamic driving force for decomposition, and the pathway with the lowest $\Delta G_P$ is the preferred decomposition pathway among the considered candidates. 
%Microscopically, this criterion reflects the competition between configurational entropy, which stabilizes chemical disorder, and enthalpic interactions, which favor chemical ordering, segregation, or phase separation in HEAs~\cite{liu2025computational}.

The enthalpy of the decomposed phases are obtained from DFT calculations of special quasi-random structures generated using ICET package and the entropy from configurational mixing entropy.
The full details are given in Supplementary material (Section {\hili VIII}, Supplementary Table {\hili S4}, and Supplementary Fig. {\hili S7}).
The temperature dependence of the Gibbs free energies of stable phases are shown in Fig.~\ref{fig:phase_separation}(a). 
The results suggest that phase decomposition of the HEA is expected to occur at around 370 K to (Mo,W)S$_2$, (Ta,Nb)S$_2$ and VS$_2$. 
%Decomposition to (MoWTaNb)S$_2$ and VS$_2$ is nearly stable at around {\hili 330} K, and 
We also see that (MoWTaNb)S$_2$ is expected to decompose to (Mo,W)S$_2$, (Ta,Nb)S$_2$ at slightly above 400 K. Overall, it appears that VS$_2$ disfavors mixing with other elements.

Next, we generated 500 quinary alloy equimolar structures on $5 \times 5$ supercells and calculated the mixing energies using the fine-tuned model, Fig.~\ref{fig:phase_separation}(b). The DFT-calculated mixing energy of the SQS (0.0269~eV/unit) and its fine-tuned model counterpart (0.0254~eV/unit) are comparable to the average value obtained from 500 random structures ($H_{\rm ave}=0.0292$~eV/unit). The width of the distribution, $\sigma = 0.0041$ eV/unit, is clearly lower than the average value. Thus, as there are no structures reaching mixing energies close to zero, the mixing at high temperatures will be purely due to the mixing entropy gain.

Divilov et al.\cite{divilov2024disordered} recently proposed a disordered enthalpy–entropy descriptor (DEED) for predicting synthesizability of HEAs.
It is defined as
\begin{equation}
    \mathrm{DEED} = \sqrt{\frac{\sigma^{-1}}{H_{\rm ave}+\Delta H_{\rm hull}}},
\end{equation}
where $\mathrm{H_{\rm ave}}$ and $\sigma$ are as defined above. $\Delta H_{\rm hull} = 0.0069$~eV/unit is the energy difference between the convex hull and the energy from the unaries (i.e., the end-points used to calculate mixing energies), and can be readily extracted from the lowest Gibbs free-energy in Fig.~\ref{fig:phase_separation}(a), here corresponding to the (Mo,W)S$_2$+(Ta,Nb)S$_2$+VS$_2$ system.
When applied to our (Mo,Ta,Nb,W,V)S$_2$ system, we obtained DEED = 82~(eV/unit)$^{-1}$ or 247~(eV/atom)$^{-1}$. 
The DEED value alone cannot be used to deduce synthesizability, as it only provides a relative tendency.
However, in Ref. \cite{divilov2024disordered}, the threshold of synthesizability was found to be about 20 (eV/atom)$^{-1}$ for transition metal carbides and carbonitrides, and about 35 (eV/atom)$^{-1}$ for borides. The value obtained here is well above these thresholds in accordance with its successful experimental synthesis\cite{cavin20212d}.

% To further verify the phase-separation result, and demonstrate the computational efficacy of the fine-tuned models, we performed Monte Carlo simulations using the MCHAMMER package in the canonical ensemble\cite{aangqvist2019icet}. As shown in Fig.~\ref{fig:phase_separation}(c), a pronounced separation of CV emerges at around 400 K. %Unlike the V–V pair, 
% Heteroatomic bonds involving V, such as Nb–V, W–V, Mo–V, and Ta–V, exhibit a continuous and significant increase in their CV values, indicating a reduction in the proportion of these bonds. In contrast, the increasing CV of the V–V bond reflects a higher fraction of V–V pairs. 
% {\hili Thus, the simultaneous increase of CV$_{\mathrm{V-V}}$ and CV$_{\mathrm{Nb-V}}$, CV$_{\mathrm{W-V}}$, CV$_{\mathrm{Mo-V}}$, and CV$_{\mathrm{Ta-V}}$ consistently indicates V segregation: V--V bonds are enriched, while V--metal heteroatomic bonds are depleted.}

% Meanwhile, other non-V-related bonds maintain CV values close to zero across all temperatures, corresponding to a nearly random {\hili distribution among Mo, Ta, Nb, and W}. These observations indicate that as the temperature decreases, VS$_2$ gradually begins to separate from the alloy at around 400 K.

To further verify the phase-separation result and demonstrate the computational efficacy of the fine-tuned models, we performed Monte Carlo simulations using the MCHAMMER package in the canonical ensemble \cite{aangqvist2019icet}. As shown in Fig.~\ref{fig:phase_separation}(c), a pronounced separation of the SRO parameters emerges at around 400 K. Heteroatomic bonds involving V, such as Nb--V, W--V, Mo--V, and Ta--V, show positive and increasing SRO values, indicating depletion of these V--metal bonds relative to random mixing. In contrast, the increasing positive SRO of the V--V bond reflects enrichment of V--V pairs. 
%Thus, the simultaneous increase of SRO$_{\mathrm{V-V}}$ and SRO$_{\mathrm{Nb-V}}$, SRO$_{\mathrm{W-V}}$, SRO$_{\mathrm{Mo-V}}$, and SRO$_{\mathrm{Ta-V}}$ consistently indicates V segregation: V--V bonds are enriched, while V--metal heteroatomic bonds are depleted.
%
Meanwhile, other non-V-related bonds maintain SRO values close to zero across all temperatures, corresponding to a nearly random distribution among Mo, Ta, Nb, and W. These observations indicate that as the temperature decreases, VS$_2$ gradually begins to separate from the alloy at around 400 K.

Interestingly, however, a partial VS$_2$ separation is already evident at much higher temperatures, something that could not have been deduced from the decomposition diagram in Fig.~\ref{fig:phase_separation}(a).
The final structural configurations at different temperatures, Fig.~\ref{fig:phase_separation}(c), also clearly show the separation.
Our results are in agreement with the experimental results of Cavin et al., who synthesized (MoTaNbWV)S$_2$ using the chemical vapor transport (CVT) method at a temperature of 1000~K\cite{cavin20212d}. 
While aiming for equimolar composition, their SEM-EDX results indicate that the five constituent elements display distinct distribution profiles. V exhibits the lowest atomic fraction, whereas Mo presents the highest atomic fraction among all measured elements. 
This is in qualitative agreement with our results, which show that V disfavors mixing with other elements, whereas Mo shows strongest tendency to mix with other elements.

\section{Conclusions}

We investigated the applicability of uMLIPs to accurately predict the energies and mixing energies of 2D high-entropy TMDC alloys. We benchmarked a few leading foundation models to three differently constructed structural databases and found all of them to yield unsatisfactory mixing energies.
We then investigated fine-tuning strategies based on random and enumerated structures.
%By systematically combining enumerated structures across different compositions and supercell sizes within a finite range, it was found that the accuracy of models gradually converge as the size of the training set increases. 
%The fine-tuned enumerated models demonstrated robust performance across different types of structures. 
Our main findings and recommendations for fine-tuning are:
\begin{enumerate}
%\item For the HEA system considered here, current foundation models were all found to exhibit insufficient accuracy. This was particularly true for the mixing energies, even when total energies seemed good.
\item Training with random structures may yield lower MAE, i.e., works better for average structure, but may also fail badly for structures beyond training set, usually the ordered ones. This makes the enumerated model safer to use in sampling the configuration space.
\item 
%More data generally yielded better model, as long as structural variety was guaranteed. This can be done systematically with the enumeration approach.
Structure generation by enumeration guarantees structural diversity, thereby leading to systematic model improvement with increasing training dataset size.
\item Reasonably good models for describing HEAs can be obtained even when trained using structures with a smaller number of alloying components. Here, e.g., model trained using 2- and 3-component alloys could be used for 5-component alloy.
\item At a large dataset size, training from scratch led to a similar MAE as from fine-tuning. However, at a smaller dataset size, fine-tuned models still result in stable structural relaxation whereas models trained from scratch showed instabilities.
\item Structure selection by Monte-Carlo annealing appeared as a good alternative to the enumeration, yielding similar model accuracy for the same dataset size. We emphasize that successful simulation of phase decomposition necessitates a model that yields accurate results for any alloy composition and any local atomic ordering.
\end{enumerate}
{\hili Table~S2 and Figure~S4 in Supplementary Materials show that these trends are very similar between MACE-small and MACE-small-0b2, thereby demonstrating that these conclusions are likely applicable to other models as well.}
We expect these findings to be applicable also to other complex alloys and material systems beside the TMDC HEAs considered in this work.
Overall, we think that the fine-tuning with enumerated structures offers an effective strategy for training a stable and accurate model.

Finally, by integrating the models with Monte Carlo simulations and random structure sampling, we analyzed the phase decomposition behavior of 5-component (MoTaNbWV)S$_2$ alloy.
Based on the Gibbs free energy diagram evaluated using SQS and on explicit Monte Carlo simulations,
we found a good agreement for decomposition temperature and for the decomposition products. 
However, the Monte Carlo simulations indicate that partial phase separation of VS$_2$ occurs already at higher temperatures. Furthermore, we evaluated the disordered enthalpy–entropy descriptor, which strongly supports the synthesizability of these alloys. Our results are in good agreement with the available experimental results.

Certainly, fine-tuned uMLIPs will not be the best solution for every situation. SQSs calculated using DFT are often good enough for initial evaluation of the phase stability, as shown also in this work. Cluster expansion (CE) is dramatically faster for evaluating the energy of a given structure, thereby allowing Monte-Carlo simulations for much larger supercells and number of steps. However, the training of CE cannot benefit from the prior chemical knowledge of uMLIPs and thus likely to require very large training dataset in the case of HEAs. Importantly, fine-tuned uMLIPs are also expected to be able to provide other material properties with fairly little additional training, such as detailed atomic structure (e.g., atom displacements from nominal lattice site), vibrational properties (e.g., IR spectrum, Raman spectrum, and vibrational contributions to free energy), mechanical properties (e.g., Young's modulus), and the interactions between the HEA and its environment, and thus highly promising for a wide range of future studies on HEAs.

\begin{acknowledgments}
The authors thank CSC–IT Center for Science Ltd. for generous grants of computer time.
We also wish to acknowledge Prof. Paul Erhart and Dr. Rico Friedrich for fruitful discussions. The authors thank Dr. Ethan Berger for his valuable support during the early stages of the project.
\end{acknowledgments}

\section*{Funding}

We are grateful to the Research Council of Finland for support under Academy Project funding No. 357483.

\section*{Competing interests}

All authors declare no financial or non-financial competing interests. 

\section*{Data availability}

The datasets and models generated and/or analysed during the current study are available in the Zenodo repository
%https://doi.org/10.5281/zenodo.17750812.
\href{https://zenodo.org/records/17750812?token=eyJhbGciOiJIUzUxMiJ9.eyJpZCI6IjI5NWQ5NTVlLTNiMjUtNGVmOS1hMzQ1LTUxNThiMDY0Mzk0ZSIsImRhdGEiOnt9LCJyYW5kb20iOiJjMDAwZDBlZTE1MTYzZWYzZTM5NmU1YTQ3M2ViNDBlZSJ9.u_ANBU56qO-nfGjwH2JP0Jy7c-iivxlzE9xynISHiLxmYFYhzuYh3fME6ZGoAPBnKbuerNpO8yq-cAGHkTGHeA}{https://doi.org/10.5281/zenodo.17750812}.

\section*{Author contributions}

H.-P.K. conceived and supervised the project. C.Z. carried out all the calculations, analyzed the results, and prepared the figures. C.Z. and H.-P.K.  wrote the main manuscript text.

%\appendix

% The \nocite command causes all entries in a bibliography to be printed out
% whether or not they are actually referenced in the text. This is appropriate
% for the sample file to show the different styles of references, but authors
% most likely will not want to use it.
%\nocite{*}

%\bibliographystyle{apsrev4-1}
%\bibliography{refs}% Produces the bibliography via BibTeX.

%apsrev4-2.bst 2019-01-14 (MD) hand-edited version of apsrev4-1.bst
%Control: key (0)
%Control: author (8) initials jnrlst
%Control: editor formatted (1) identically to author
%Control: production of article title (0) allowed
%Control: page (0) single
%Control: year (1) truncated
%Control: production of eprint (0) enabled
%

\end{document}